\newcommand{\be}{\begin{equation}}
\newcommand{\ee}{\end{equation}}
\newcommand{\bea}{\begin{eqnarray}}
\newcommand{\eea}{\end{eqnarray}}
\newcommand{\ba}[1]{\begin{array}{#1}}
\newcommand{\ea}{\end{array}}

\newcommand{\mb}{\mathbf}

\newcommand{\lan}{\langle}
\newcommand{\ran}{\rangle}

\newcommand{\diracslash}[1]{#1\llap{/\kern2pt}}

\documentclass[aps,prl,preprint,amsmath,showpacs,amssymb]{revtex4}
\usepackage{epsfig}
\usepackage{amssymb}
\usepackage{times}
\usepackage{amssymb}
\usepackage{amsmath}
\usepackage{mathrsfs}
\usepackage{graphicx}
\usepackage{epsfig}
\usepackage{dcolumn}
\usepackage{color}
\usepackage{bm}
\begin{document}
\setlength{\topmargin}{0.2in}

\title{Quantum Effects at Low Energy Atom-Molecule Interface }
%\title{Quantum Physics and Chemistry with Cold Atoms and Molecules at
%Atom-Molecule Interface}
\author{B. Deb$^{1,2}$, A. Rakshit$^1$, J. Hazra$^1$ and D. Chakraborty$^1$}
\address{$^1$ Department of Materials Science, $^2$ Raman Centre for Atomic, Molecular and Optical Sciences (RCAMOS),
 Indian Association for the Cultivation of Science,
 Jadavpur, Kolkata 700032. INDIA}

\begin{abstract}
Quantum interference effects in inter-conversion between cold atoms and diatomic molecules are analysed. Within the framework of Fano's theory, continuum-bound anisotropic dressed state formalism  of atom-molecule quantum dynamics is presented. This formalism is applicable in photo- and magneto-associative
strong-coupling regimes. The significance of Fano effect in ultracold atom-molecule  transitions is discussed.  Quantum effects at low energy atom-molecule interface are important for exploring  coherent phenomena in hither-to unexplored parameter regimes. \\
\\
{\bf Keywords} : Continuum-bound dressed state, Atom-molecule coherence,  Photoassociation, Magnetic Feshbach resonance, Optical Feshbach resonance,  Quantum interference
\end{abstract}

\pacs{32.80.Qk,34.50.Rk,03.65.Nk,34.10.+x}

\maketitle

\section{1. introduction}

Ever since the realisation of Bose-Einstein condensation \cite{science95anderson,hulet95prl,ketterle95prl} in dilute gases of alkali atoms  sixteen years back, there has been a revolutionary growth
in research activities with cold and ultracold atoms.  By `cold atoms'  we mean  a temperature range for atoms from   miliKelvin (mK) down to microKelvin
($\mu$K) regimes  while `ultracold atoms' imply atoms in the  sub-$\mu$K or  nanoKelvin (nK) temperature regime. Thanks to the tremendous development in the technology of cooling and trapping
of atoms \cite{rmp98cohen,rmp98chu,rmp98phillips} in the 80's and early 90's culminating in the first demonstrations of Bose-Einstein condensation  \cite{science95anderson,hulet95prl,ketterle95prl}.  1997 Nobel prize in physics was jointly awarded to Steven Chu, William D. Phillips and Claude N. Cohen-Tannoudji for their contributions to laser cooling. Carl E. Wieman, Eric A. Cornell and Wolfgang Ketterle were awarded 2001 Nobel prize in physics for the first achievement of Bose-Einstein condensate (BEC) in dilute atomic gases.  Over the years,
several research groups around the world have demonstrated Bose-Einstein condensation in different alkali and other kinds of atomic gases. Along with the progress of research in cold bosonic atoms,  there has been a lot of advancement in cooling and trapping of fermionic atoms.  Fermi degeneracy in a Fermi gas of alkali atoms was first demonstrated by Deborah Jin in 1999 \cite{jin99prl}.

Cold atoms have revitalised not only atomic physics, but many other  areas of research starting from superfluidity, molecular and condensed matter physics to astro-chemistry.
This has been possible because of extraordinary properties of cold atoms making them a fertile  ground for studying new physics and chemistry. One of the most interesting properties of cold atoms is the tunability of interatomic interaction  over a wide range  by an external magnetic field Feshbach resonance (MFR) \cite{pra93verhaar, rmp2006kohler, rmp2010chin}.
 By changing the strength of an external magnetic field near a Feshbach resonance, one can basically alter $s$-wave scattering length of ultracold atoms from large positive to large negative values or vice versa. Exactly at resonance, the scattering length diverges. Near Feshbach resonance,
 an atomic gas becomes strongly interacting. Over the years, MFR has been extensively used as a standard tool for studying strongly interacting atomic Fermi gases \cite{science2002thomas, rmp2008bloch,rmp2008stringari}. In fact, it has facilitated to achieve a number of milestones in the area of Fermi gases : (1) realization of  $s$-wave fermionic superfluidity \cite{nature2005ketterle}
 in a Fermi gas of alkali atoms, (2) achievement of molecular Bose-Einstein condensation \cite{nature2003jin,science2003jochim,prl2003martin}
 of a new type of diatomic molecules known as `Feshbach' molecules formed
from cold fermionic atoms and (3) demonstration  \cite{nature2005ketterle,prl2007grim,prl2007ketterle} of a crossover between Bardeen-Cooper-Schrieffer (BCS) state and BEC known as BEC-BCS or BCS-BEC crossover \cite{jltp1985noziers,prl1993randeria}. All these progresses  along with the development of laser-generated artificial crystalline structures for cold atoms known as optical lattices  have made cold atoms a test bed for the models of condensed matter physics. One of the advantages for simulating fundamental
quantum phenomena of condensed matter systems with cold or ultracold atoms is that, unlike electrons in solid state materials,  cold atoms allow unprecedented control over atom-atom interaction. New insight into the quantum physics of many-body systems can be developed to enrich
our understanding on unsolved problems in physics such as high temperature superconductivity \cite{htsup}. In one hand  cold atom science offers new opportunities
for quantum simulation of condensed matter phenomena, on the other hand research into quantum optical phenomena with interacting cold or ultracold atoms is opening up new vistas of
 research into hither-to unexplored quantum effects at the interface of atomic and molecular states. The advent of Feshbach molecules due to MFR and the formation of cold molecules by
 photoassociation (PA) \cite{prl87thorsheim,rmp1999weiner,rmp2006jones} have generated a lot of research interest in physics as well as in chemistry. Both the methods of PA and MFR have  several common features. An analogy of PA with MFR has been exploited to develop an optical method of altering $s$-wave scattering length known as optical Feshbach resonance (OFR) \cite{prl:1996:fedichev,pra:1997:bohn,prl:2002:fatemi}. Although OFR has been so far found to be not as effective as MFR as far as tuning of $s$-wave scattering length is concerned, optical methods based on strong-coupling PA with quantum interference between optical transition pathways has been theoretically shown to be useful for altering higher partial-wave atom-atom interaction \cite{debarxiv}.

In this review article we present an overview on coherent phenomena involving atom-molecule transitions at low energy with a particular focus on  quantum interference
effects at atom-molecule interface (AMI).  It is almost impossible to make such a review
self-content  given the diversity  of current state-of-art research with ultracold atoms and molecules.
There are  numerous  excellent reviews, monograms and books  on cooling and trapping of atoms, precision spectroscopy, Bose-Einstein condensates and related subjects written over the
last one and a half decade. It is not our aim to give an overview of these topical subjects that are well documented in the literature.  Instead, we refer to some reviews and books
\cite{becreviews,metcalfbook,bookbecpethik,bookLaserCoolingLetokhov} for general readers. Our discussions are mainly centred around some selective theoretical topics related to inter-conversion between atoms and  molecules at ultra low energy   in the
presence of external fields.  Ultracold molecules are an emerging area of research with tremendous prospects for exploring new physics and chemistry.
 In recent times a few focus or special articles have appeared in scientific magazines and journals revealing the broad and general interest of research with ultracold molecules. For instance, we can cite the recent article by Jin and Ye \cite{phystoday2011Jin}, the review  by  Carr {\it et al.} \cite{njp2009ye} and the book  on cold molecules \cite{kremsbook}.

 In  section 2, after a brief discussion on the formation of molecules from cold or ultracold atoms, continuum-bound spectroscopy is described with an emphasis on
the underlying processes of association of two atoms into a molecule.
The processes discussed are photoassociation (PA) and magnetic Feshbach resonance (MFR). Section 2 provides
a general discussion on the associative processes  while the technical aspects of PA and MFR are described in sections 3 and 4.
In section 3  continuum-bound dressed state formalism is presented for describing quantum dynamics at AMI.  Section 4  is primarily devoted to the analysis of quantum interference phenomena that can arise
in atom-molecule transitions.
In case of  multiple  pathways of dipole transitions at AMI, quantum interference
 may arise between two competing pathways. These interference effects can   be controlled by an external static or dynamic magnetic or electric field or combination of several fields.  In section 5,  conclusion and  a brief  discussion on  prospects of new physics and chemistry at  AMI.

\section{2. Continuum-bound spectroscopy: Photoassociation and Feshbach resonance}

Photodissociation of diatomic molecules is a well known bound-continuum process where a molecule is fragmented into
its constituent atoms due to absorption of light. Photodissociation spectroscopy has been long studied and developed
into a standard method of bound-continuum or continuum-bound spectroscopy. In contrast, photoassociation  is a relatively new kind of continuum-bound spectroscopy developed mostly  over the last 25 years in the modern era of laser cooling and trapping of atoms much below 1 K. Photoassociation is a process by which two colliding ground state cold atoms in the presence of
laser light of appropriate frequency can be associated to form an excited diatomic molecule via absorption of one photon. So,this is a dipole transition from the  continuum (two unbound ground state atoms in the state of collision) of ground diatomic molecular potential to the ro-vibrational bound states in the electronically excited molecular potential. Excited molecule formed by photoassociation is  long-ranged compared to the typical range of  diatomic molecules ($\le$ 20 a$_0$, where $a_0$ is the Bohr radius) at room temperature. The binding energies of excited molecules formed by photoassociation lie very close to  the dissociation threshold of the molecular potential - typically in the range of sub-mK to several
hundred mK below the threshold. Therefore, the initial temperature of the atomic cloud should be cold enough to
access such loosely bound molecular states by photoassociation. While dissociation of a diatomic molecule by
an external perturbation is possible at room temperature, associating two colliding atoms into a diatomic bound state requires cold temperature because of the nature of molecular continuum-bound spectroscopy that we will discuss shortly. This is why associating spectroscopy due to an external field has not been developed into a full flagged discipline until the advent of cold atoms.

\begin{figure}
\includegraphics[width=3.5in]{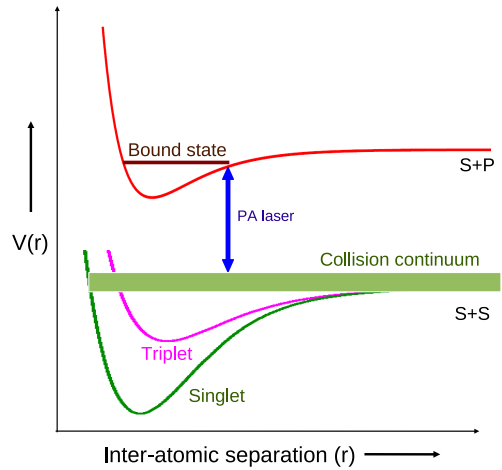}
\caption{A schematic diagram of photoassociation (PA). Shown are  ground and excited state adiabatic molecular potentials.
Singlet and triplet ground state potentials are mixed up due to hyperfine interaction. PA laser excites two ground state ($S$) atoms colliding in the continuum of molecular ground state potential into a bound state in the excited molecular potential which asymptotically corresponds to one ground ($S$) and another excited ($P$) atom. }
 \label{fig1}
  \end{figure}

Not only PA spectroscopy,  advancement  in cold atom research has led to another new kind of continuum-bound spectroscopy due to magnetic field Feshbach resonance (MFR). While PA is due to electric dipole continuum-bound transitions involving excited electronic configurations, MFR spectroscopy is based on magnetic coupling between continuum and a quasi-bound state at near-zero collision energy  with quasi-bound and continuum states belonging to different  configurations (scattering channels) of two ground state atoms. MFR spectroscopy has led to the emergence of Feshbach molecules which are produced due to three-body collisions near magnetic field at which two-body scattering (Feshbach) resonance occurs. These molecules are translationally and rotationally cold. Since their binding energies are close to the threshold of ground molecular potentials, they have extremely high vibrational quantum numbers. If a Feshbach molecule is formed from a pair of two-component fermionic atoms,
the molecule attains extraordinary stability, because inelastic molecule-atom or molecule-molecule scattering
is suppressed due to Pauli blocking. With magnetically tunable Feshbach resonance method,
two-component Fermi gas of ultracold atoms are convertible into Bose-Einstein
condensates of Feshbach molecules.

The experimental study of PA in the presence of MFR  was initiated in late 90's \cite{prl1998heinzen,pra1998verhaar,prl1999vladan,prl2003chu, pra2004chu}.
Over the last 4 years, several such experiments \cite{prl2008hulet,prl2007winkler,science2008ni,nphys2009rempe,pra2009rempe} have been
carried out. The results of these experiments as well as theoretical studies \cite{prl2008mackie,njp2008cote,prl2008cote,njp2009cote,jpb2009bdeb,jpb2009agarwal,jpb2010bdeb} reveal significant effects of magnetic field induced  Feshbach resonances on PA  and cold collision properties.  According to Franck-Condon principle, the  intensity
of bound-bound (molecule-molecule) or continuum-bound (dissociative or associative)  spectrum is in general proportional to
the square of overlap integral between the two bound states or between continuum and bound state. Therefore,  a continuum-bound
transition is most probable when the most prominent (with the largest amplitude) antinodes of  continuum and
bound states are located at interatomic separations which are nearly equal or comparable. When the antinode of either continuum
or bound state lies at a separation close to the node of the other, the probability of transition between such continuum and bound state
is the least. In the presence of Feshbach resonance, enhancement and suppression in PA spectral intensity profile can take
place due to quantum interference between photoassociative and Feshbach resonances leading to Fano-type asymmetric spectral profile.
Such quantum interference at AMI has important implications in coherent control of atom-atom cold collision
and bound state properties. For instance, quantum interference in the strong PA coupling limit leads
to nonlinear Fano effect and suppression of power broadening  in PA spectrum, double resonances in atom-atom
scattering and manipulation of higher partial-wave interaction. Studies on quantum interference at AMI may
lead to  another new kind of continuum-bound spectroscopy which we may call as Fano-Feshbach quantum interference spectroscopy (FFQIS).

PA spectrum is usually determined by measuring the loss of atoms from atomic traps due to a PA laser. Once a bound state is formed
in an excited molecular potential by PA process, this bound state can spontaneously decay to either two hot atoms or a bound state in
ground molecular potential close to dissociation threshold. These hot atoms or ground molecules in highly excited vibrational levels then escape
from the trap. Theoretical treatment  of PA spectrum
views PA process as laser-assisted or laser modified atom-atom scattering with spontaneous decay from excited bound states.  Bohn and Julienne \cite{pra96bohn, pra99bohn}
 have developed a
semi-analytical theory of PA spectroscopy based on multi-channel scattering theory with an artificial lossy channel. The characteristic features of PA spectroscopy as trap loss
can be drastically changed mainly in three situations: (1) Strong-coupling photoassociative regime, that is, a regime where
photoassociative dipole coupling exceeds spontaneous emission line width (2) Interference of MFR with PA and (3) The application of
multiple lasers driving multiple continuum-bound dipole transitions in the strong-coupling regime.

In strong photoassociative dipole coupling \cite{prl2009jisha,jishathesis}, rate of stimulated bound-continuum transition can take over the rate of spontaneous transition recycling the atoms into the initial continuum and thereby  stalling trap loss.  As a result,
atom-molecule coherence between the ground continuum and the excited molecular state can emerge forming a continuum-bound ``dressed state". It is then appropriate
 to describe  PA dynamics in the framework of a dressed state picture. Fifty years back, in a classic paper Fano \cite{physrev1961fano} showed  how   to diagonalise  an
  interacting continuum-bound  system  resulting in
  ``dressed continuum". In late 50's, Feshbach \cite{annphys1958feshbach} formulated the mathematical theory of a particular type of two-body scattering resonances now known as
   ``Feshbach resonance" in literature. For Feshbach  resonance to occur,  three physical conditions need to be satisfied:
   (i) existence of a quasi-bound state supported by a closed scattering channel is needed,
   (ii)  a coupling between this closed channel and open channel or channels s is required and (iii) initial collision energy has to be close to the binding energy of
   the quasi-bound state. In multichannel two-body scattering theory, a channel is defined by a possible arrangement of  internal degrees-of-freedom (
   such as electronic  or hyperfine spin, angular momentum of relative motion etc.) in the asymptotically large separation limit of two bodies. In essence, both
   Fano and Feshbach methods deal with a continuum interacting with one (more than one)  bound state (states). Therefore, both methods are related, although
  the formalism and the physical contexts in which both methods were originally developed  are quite different. In the absence of any loss or inelastic process,
  both methods provide exact treatments  of an interacting continuum-bound system and therefore they are valid in any coupling regime.    PA  is usually viewed as a trap loss spectroscopy studied mostly in the weak-coupling regime, although PA theory has
  made use of Fano's method \cite{pra96bohn,pra99bohn}.  Description of strong-coupling PA requires  an explicit dressed state picture based on Fano-Feshbach diagonalisation method. There are several interesting strong-coupling effects which include (a) large shifts and line width exceeding spontaneous
  line width, (b) occurrence of multi-photon process, (c) generation of higher partial-wave interaction in the continuum and (d) excitations of higher
  rotational levels in excited molecules.

When PA occurs in the presence of MFR, the existence of a closed channel quasi-bound state opens up a bound-bound dipole  transition pathway
  due to PA laser. In such a physical situation, there are three competing couplings of which two are continuum-bound  and one
  is the bound-bound type. Of the two continuum-bound couplings, one is PA coupling and the other is magnetic one between the
  closed channel bound state and the ground scattering state.  Naturally,  quantum interference arises between any two of these three possible transition
  pathways. Strong-coupling phenomena are then largely influenced by these quantum interference effects. Theoretical treatment of quantum interference in
  strong-coupling regime of PA rests on the method of Fano-Feshbach diagonalisation resulting in dressed continuum. In what follows we  show that
  quantum interference at AMI can significantly manipulate molecule formation and continuum states. In section III, we  present the basic formalism
  for obtaining dressed continuum that is central to the description of quantum interference effects at AMI as discussed in section IV.

\section{3. Dressed state formalism of continuum-bound quantum dynamics}

As mentioned in the preceding section, following  Fano or Feshbach method one can diagonalise an interacting system of
continuum and bound states to obtain dressed continuum.
The method we follow to diagonalise a PA system is partly a synthesis of the two methods. Our method
is based on real space Green functions that are used to solve second order coupled differential equations.
Since our interest lies in the manipulation of bound and scattering states by strong-coupling effects, we
prefer to work in coordinate space to derive explicitly scattering amplitudes that can be determined
from asymptotic analysis of dressed continuum at large separations. Let us consider PA process as schematically
depicted in Fig.1. To begin with, let us idealise PA system by assuming that the excited bound state is loss less,
that is, it has infinite life time. Spontaneous decay of the excited state will be introduced later.
Let the PA laser be tuned near resonance of a particular ro-vibrational
bound state characterised by vibrational quantum number $v$ and rotational quantum number $J$. Generally, $J$ is given
by $\mathbf{J} = \mathbf{L} + \mathbf{S} + \vec{\ell}$, where $\mathbf{L} = \mathbf{l}_a + \mb{l}_b$ is the total
electronic orbital angular momentum of the two atom with $\mb{l}_{\alpha}$ being the atomic electronic orbital angular momentum of $\alpha(a,b)$-atom,  $\mathbf{S}$ is the total electronic spin angular momentum and $\vec{\ell}$ represents
the angular momentum of the relative motion of the two atoms. In case of bound-bound spectroscopy of diatomic molecule,
angular momentum of relative motion or rotation of internuclear axis is usually represented by the symbol `$N$' instead
of $\ell$.

 For simplicity,
let us assume that the two ground state atoms collide in a single channel (configuration of internal degrees-of-freedom
of the two separated atoms) denoted by $\mid g \rangle \equiv \mid f_a,f_b, F \rangle$, where $f_{a(b)}$ is the hyperfine quantum number of atom $a (b)$ and
$\mathbf{F} = \mathbf{f}_a + \mathbf{f}_b$. Let the angular state of the excited bound state
be denoted by $\mid e \rangle $ which includes all possible angular quantum numbers to characterise the excited state
including the rotational quantum number $J$ and its projection  $\zeta$  on
the body-fixed $z$-axis which is the internuclear axis. Under Born-Oppenheimer approximation, integrating over electronic coordinates, the Hamiltonian of the system can be written as $H = H_{\rm{rel}} + H_{\rm{CM}}$, where $H_{\rm{CM}} = -\frac{\hbar^2 }{2 M} \nabla_R^2$ describes centre-of-mass motion of the two atoms with total mass $M = m_{a} + m_{b}$ where $m_{a(b)}$ is the mass of atoms $a (b)$, $\nabla_R$ represents Laplacian corresponding to the centre-of-mass position vector  $\mathbf{R} = (m_a \mathbf{r}_a + m_b \mathbf{r}_b)/M$ with $\mathbf{r}_{a(b)}$ being the position vector of atom $a(b)$. The Hamiltonian of relative motion the two atoms is given by
\bea
H_{\rm{rel}} = - \frac{\hbar^2}{2 \mu } \nabla_r^2 + \sum_{\alpha = g,e}\left (E_{\alpha} + V_{\alpha} \right ) \mid \alpha \rangle \langle \alpha \mid
+ \left ( e^{- i \omega_L t} V_{int}^{(\rm{PA})} \mid e \rangle \langle g \mid + \rm{C.c.}  \right )
\eea
where $\nabla_r$ is the Laplacian corresponding to the relative position vector
$\mathbf{r} = \mathbf{r}_a - \mathbf{r}_b$, $\mu = m_a m_b/M$ is the reduced mass and $ \omega_L $ is the angular frequency
of  PA laser. Here $E_{\alpha}$ is the asymptotic internal (electronic) energy of the molecular state $\mid \alpha \ran$, that is, total
 internal energy of the two atoms at large separation (separated atom limit).
 $V_{\alpha}$ denotes
the atom-atom interaction in state $\alpha = g,e$ and $V_{int}^{(\rm{PA})} = -\vec{\mathscr{D}} \cdot \vec{{\cal E}}$ is
the PA laser induced dipole interaction where $\vec{\mathscr{D}}$ is the molecular transition dipole moment and $\vec{{\cal E}}$ is the electric
field of PA laser. The ground molecular state ($\alpha=g$) at large separation corresponds to two $S$ ($l_a = l_b  =0$) atoms and hence
we can set $E_g=0$ while the excited state ($\alpha=g$) in the separated atom limit corresponds to one $S$ and another $P$ (either $l_a$ or $l_b $ being equal to 1) and so $E_e = \hbar \omega_A$ where $\omega_A$ is the atomic transition frequency between $S$ and $P$ states.
Since
$H_{\rm{CM}}$ describes only free motion of the centre-of-mass of the two atoms, the wave function is separable in
two parts -  centre-of-mass and relative wave functions. The centre-of-mass motion does not at all influence
atom-atom scattering and  introduces only a  phase factor
$\exp[i \mathbf{K}\cdot \mathbf{R}]$  to the total wave function. We therefore henceforth discuss only relative motion of the two atoms.

The time-dependent relative wave function can be formally expressed as $\mid \Psi(t) \rangle = e^{-i H_{\rm{rel}} t/\hbar} \mid \Psi(t=0) \rangle $. Suppose, the Hamiltonian $H_{\rm{rel}}$ satisfies the eigenvalue equation
$H_{\rm{rel}} \mid E \rangle_{\rm{dr}} = E \mid E \rangle_{\rm{dr}}$ where
$\mid E \rangle_{\rm{dr}}$ is  the eigen function known as dressed continuum with $E$ being the eigen energy.
These dressed basis are energy-normalised, that is,
$_{\rm{dr}}\langle E' \mid E \rangle_{\rm{dr}} = \delta(E-E')$. One then defines an identity operator $\hat{\mb{I}} = \int dE \mid E \rangle_{\rm{dr}} \, _{dr}\lan E \mid $. Making use of this identity operator,
 $\mid \Psi(t) \rangle$ can be expanded in terms of these dressed continuum basis in the following form
\bea \mid \Psi(t) \rangle = \int d E e^{-iEt/\hbar} \mid E \rangle_{\rm{dr}} \, _{\rm{dr}}\lan E \mid \Psi(t=0)  \rangle
\eea
 Let the rotational energy spacing of excited ro-vibrational states are much larger than PA laser
line width such that the laser can effectively couple only a single rotational level.
In the absence of any external magnetic field,  the excited ro-vibrational level  has $(2J+1)$ fold degeneracy. We therefore need
to consider dressed state of $(2J+1)$ degenerate bound states interacting with a single continuum. The dressed continuum for such an interacting  system can be expressed as
\bea
\mid E \rangle_{\rm{dr}} = \sum_{M} A_{JM}(E, \hat{k}) \mid b \rangle \mid e(J\zeta M) \rangle + \sum_{\ell m_{\ell}} \int d E'  C_{E',\ell m_{\ell}}(E,\hat{k}) \mid E' \ell m_{\ell} \rangle_{\rm{bare}} \mid g \rangle
\eea
where $M$
is the projection of $J$ on the space-fixed (laboratory frame) $z$-axis, $\hat{k}$  denotes the direction
of the incident and scattered relative momentum;  $\mid b \rangle $ and $\mid E' \rangle_{\rm{bare}}$ are the bare bound (ro-vibrational) and bare continuum state, respectively. $A_{JM}(E,\hat{k})$ and  $C_{E'}(E, \hat{k})$ are the laser-dependent coefficients to be determined. $A_{JM}(E, \hat{k})$  can be expanded in the form
 $A_{JM}(E, \hat{k}) = \sum_{\ell' m_{\ell'}} A_{JM}^{\ell' m_{\ell'}} Y_{\ell' m_{\ell'}}^*(\hat{k})$ and similarly $C_{E',\ell m_{\ell}}(E,\hat{k}) = \sum_{\ell' m_{\ell'}} C_{E',\ell m_{\ell}}^{\ell' m_{\ell'}} Y_{\ell' m_{\ell'}}^*(\hat{k})$.

The coordinate
representation of the dressed continuum  is $\langle \mathbf{r} \mid  E \rangle_{\rm{dr}} = \Psi_{\hat{k},E}(\mathbf{r}) $
and that of bare bound state is
$\langle \mathbf{r} \mid b \rangle \mid e(J\zeta M) \rangle = \langle r \mid b \rangle \langle \hat{r} \mid e(J\zeta)  \rangle =
r^{-1}  \phi_{vJ}(r)  \mid J\zeta M; \eta \ran $ where
\bea \mid J\zeta M; \eta \ran =   i^J \sqrt{\frac{2 J + 1}{8\pi^2}} {\cal D}^{(J)}_{M \zeta }(\hat{r}) \mid \eta \ran. \eea
Here
 ${\cal D}^{J}_{M \zeta}(\hat{r}) $ is Wigner rotation matrix element and
$\hat{r}$ represents Euler angles and $\mid \eta \ran$ denotes the electronic or hyperfine spin  angular  state of the excited bound level. Because of the cylindrical symmetry of diatomic molecules with respect
to internuclear axis, $\zeta$ is a good quantum number. Let us assume that the excited molecular state
belongs to Hund's case a or b. Then $\Lambda$ which is the projection of $L$ on internuclear axis is also
a good quantum number. For simplicity,  we consider that the excited state has $\Sigma$ symmetry, that is $\Lambda = 0$,  and $S = 0$ (spin singlet).
 For $\zeta=0$,   $ \mid J\zeta M; \eta \ran$ reduces to  $Y_{JM}(\hat{r})\mid \eta \ran$ with $Y_{JM}(\hat{r})$ being the  spherical
harmonics. Like dressed continuum, the bare continuum states are also energy-normalised. In the cold
collision regime (see Appendix-A), the lowest partial wave $\ell=0$ ($s$-wave) is the most significant one and a few higher
partial waves (such as $p$-, $d$-wave and so on) may have finite contributions depending on the temperature of the
cold atomic cloud. It is  advantageous to expand the coordinate representation of bare continuum
$\lan \mb{r} \mid E',\ell m_{\ell} \ran_{\rm{bare}} = \frac{1}{r} \psi_{E, \ell m_{\ell}}(r) Y_{\ell m_{\ell}}(\hat{r})$
where  the wave function $\psi_{E \ell m_{\ell}}(r)$ satisfies the time-independent Schroedinger equation
\bea
\left [ \frac{\hbar^2}{2 \mu} \left \{ - \frac{d^2}{dr^2}  + \frac{\ell(\ell+1)}{r^2} \right \} + V_g(r) \right ]
\psi_{E, \ell m_{\ell}}(r) = E \psi_{E, \ell m_{\ell}}(r)
\eea
If the ground potential $V_g(r)$ is spherically symmetric, then $\psi_{E \ell m_{\ell}}(r)$ becomes independent of
$m_{\ell}$. Note that, although bare continuum can be isotropic in case of spherically symmetric ground state
atom-atom potentials, dressed continuum becomes aniosotropic due to PA coupling  which is
essentially anisotropic. In fact, any external field modified scattering is anisotropic in general.
Therefore, we need to incorporate the mathematical formulation of anisotropic scattering within the framework of  Fano's theory for a
description of strong-coupling PA. Using the coordinate representations of bare excited bound and ground continuum states we
can express the dressed continuum of Eq.(3) in the following form
\bea
\Psi_E(\hat{k},\mb{r}) = \frac{1}{r} \left [ \sum_{M} \Phi_{E,vJ}(\hat{k}, r)  \mid J\zeta M; \eta \ran +
\int d E' \sum_{\ell m_{\ell}} \Psi_{E',\ell m_{\ell}}(\hat{k}, E,r)  \mid \ell 0 m_{\ell}; g \ran \right ]
\eea
where $\Phi_{E,vJ}(\hat{k},r) = \sum_{\ell' m_{\ell'}} A_{JM}^{\ell' m_{\ell'}}(E) \phi_{vJ}(r) Y_{\ell' m_{\ell'}}(\hat{k})$,
\bea \Psi_{E',\ell m_{\ell}}(E,\hat{k}, r) = \sum_{\ell' m_{\ell'}} C_{E' \ell m_{\ell}}^{\ell' m_{\ell'}}(E) \psi_{E', \ell m_{\ell}}(r) Y_{\ell' m_{\ell'}}(\hat{k}) \eea and
$\mid \ell 0 m_{\ell}; g \ran  = \mid \ell 0 m_{\ell}\ran \mid g \ran = Y_{\ell m_{\ell}}(\hat{r}) \mid g \ran$. In what
 follows we calculate
the coefficients $A_{JM}^{\ell' m_{\ell'}}$ and $C_{E',\ell m_{\ell}}^{\ell' m_{\ell'}}(E)$ by solving the two coupled differential equations
by the method of  Green function.

Substituting Eq.(6) in time-independent Schroedinger equation $H_{\rm{rel}} \Psi_E(\mb{r}) = E \Psi_E(\mb{r})$, we obtain the following two coupled differential equations
\bea
\left [ -\frac{\hbar^2}{2\mu}\frac{d^2}{d r^2} + B_J(r) + V_{e}(r) - \hbar \delta_L - E  \right ]  \Phi_{E,v J}
 = - \sum_{\ell m_{\ell}} V_{JM;\ell m_{\ell}}^{\rm{PA}}(r) \tilde{\Psi}_{E, \ell m_{\ell}}
\label{coup1}\eea
\bea
\left [ - \frac{\hbar^2}{2\mu}\frac{d^2}{d r^2} + B_{\ell}(r) + V_{g}(r) - E \right]\tilde{\Psi}_{E, \ell m_{\ell}}
= -\sum_{M}  V_{\ell m_{\ell}; J M}^{\rm{PA}}(r) \Phi_{E,v J}
\label{coup2}
\eea
where $\delta_L = \omega_L - \omega_A$ is the laser-atom detuning,
 $B_J(r) = \hbar^2/(2\mu r^2) J(J+1)$ is the
rotational term of excited molecular bound state,  $B_\ell(r) =
\hbar^2/(2\mu r^2) \ell(\ell +1)$ is the centrifugal term in
collision of two ground state  atoms and  $V_{JM;\ell m_{\ell}}^{\rm{PA}} = \lan \ell 0 m_{\ell}; g \mid V_{int}^{\rm{PA}} \mid J\zeta M; \eta \ran $ is the electronic coupling matrix element for the transition
  $\mid J\zeta M; \eta \ran \rightarrow \mid \ell 0 m_{\ell}; g \ran$.   Here
$\tilde{\Psi}_{E \ell m_{\ell}} =  \int dE' \Psi_{E',\ell m_{\ell}}(E,r) = \int dE' \sum_{\ell' m_{\ell'}} C_{E',\ell m_{\ell}}^{\ell' m_{\ell'}}(E) \psi_{E',\ell m_{\ell}}(r)$. The coupling $V_{JM;\ell m_{\ell}}^{\rm{PA}}$ depends weakly on internuclear separation $r$.

The two coupled equations (\ref{coup1}) and   (\ref{coup2}) reduce to two separated homogeneous equations in the absence of PA laser ($V_{int}^{\rm{PA}}=0$). Therefore, the coupled equations  are exactly solvable by the techniques of Green function provided the solutions of the corresponding homogeneous equations  are known.
Let the bound state solution of the homogeneous part
of (\ref{coup1}) with  energy $E_{vJ}$ be represented by $\phi_{vJ}(r)$.
 The corresponding Green function is  \be G_{v}(r,r')= -
\frac{\phi_{vJ}(r)\phi_{vJ}(r')}{\hbar \delta_L + E - E_{vJ}} \ee
The Green
function for the homogeneous part of (\ref{coup2}) can be expressed as
 \bea
{\cal K}_{E, \ell} (r,r') = - \pi [\psi_{E \ell}^{reg}(r)\psi_{E \ell}^{irr}(r') + i \psi_{E \ell }^{reg}(r)\psi_{E \ell }^{reg}(r')], \hspace{0.5cm} r' > r \label{green1} \eea
\bea
{\cal K}_{E, \ell} (r,r') = - \pi [\psi_{E \ell}^{ reg}(r')\psi_{E \ell}^{ irr}(r)+i\psi_{E \ell}^{reg}(r)\psi_{E \ell}^{ reg}(r')], \hspace{0.5cm} r' < r  \label{green2}  \eea
Where $\psi_{E \ell}^{
reg}(r)$ and $\psi_{E \ell}^{ irr}(r)$ represent the regular and
irregular scattering solutions  in the absence of laser field.
$\psi_{E \ell}^{ reg}(r)$ goes to zero at $r = 0 $ while  $\psi_{E
\ell}^{irr}(r)$ is defined by boundary condition  at
$r\rightarrow\infty$ only. Following the preocedure as described in Ref.\cite{prl2009jisha}, we can expand
$A_{JM}(E) = \sum_{\ell m_{\ell}}  A_{JM}^{\ell m_{\ell}}$ with
\bea A_{JM}^{\ell m_{\ell}}(E) = \frac{
\Lambda_{J M}^{\ell m_{\ell}}(E)}{\hbar \delta_L + E - (E_{vJ} +
E_{vJ}^{shift}) + i \hbar \Gamma_{JM}(E)/2 } \eea
where
\bea \Lambda_{J M}^{\ell m_{\ell}}(E) =
\int \phi_{vJ}(r)
 V_{J M, \ell m_{\ell}}^{\rm{PA}}(r) \psi_{E,\ell m_{\ell}}(r) d r
\eea
is the continuum-bound molecular dipole matrix element. Here
$ \Gamma_{JM} =
\sum_{\ell m_{\ell}} \Gamma_{JM,\ell m_{\ell}}$ with  \bea \Gamma_{JM,\ell m_{\ell}}(E) = \frac{2 \pi}{\hbar} \left |  \Lambda_{J M}^{\ell m_{\ell}}(E) \right |^2 \eea being the bound-continuum partial stimulated line width. The total shift $E_{JM}^{shift} =
\sum_{\ell m_{\ell}} E_{JM,\ell m_{\ell}}^{shift}$ where the partial shift is given by
\bea
E_{JM, \ell m_{\ell} }^{{\mathrm shift}} = \int\int d r' d r
\phi_{vJ}(r') V_{J M, \ell m_{\ell}}^{\rm{PA}} (r'){\rm Re}[{\cal K_{\ell}}(r',r)]
   V_{\ell m_{\ell}, J M}^{\rm{PA}}(r)\phi_{vJ}(r).
\label{shift} \eea
The coefficient  $A_{JM}^{\ell m_{\ell}}(E)$ can be expressed in a compact form as
\bea
A_{JM}^{\ell m_{\ell}}(E) = \left [ \frac{
\Lambda_{J M}^{\ell m_{\ell}}}{ \hbar \Gamma_{JM}(E)/2} \right ] \frac{1}{\epsilon_L + i} \eea
where $\epsilon_L = [E + \hbar \delta_L - (E_{vJ} +
E_{vJ}^{shift}]/ (\hbar \Gamma_{JM}(E)/2) $. Thus, with explicit analytical form of $A_{JM}^{\ell m_{\ell}}(E)$ we have $\Phi_{E,vJ}(r) = \sum_{\ell m_{\ell}} A_{JM}^{\ell m_{\ell}}(E)\phi_{vJ}(r)$. Substituting this into Eq.(\ref{coup2}), and making use of Green function of Eqs. (\ref{green1}) and  (\ref{green2}), we obtain
\bea  \tilde{\Psi}_{E',\ell m_{\ell}}(E,r) &=&  \psi_{E,\ell m_{\ell}}^{\rm{reg}}(r) - \sum_{\ell' m_{\ell'}} \int dE'\sum_{M} A_{JM}^{\ell' m_{\ell'}}(E) \nonumber \\
&\times& \int dr' {\cal K}_{E',\ell m_{\ell}}(r,r')  V_{\ell m_{\ell}; J M}^{\rm{PA}}(r') \phi_{vJ}(r'). \label{dss} \eea
Now, comparing the above expression with $ \tilde{\Psi}_{E',\ell m_{\ell}}(E,r) = \sum_{\ell' m_{\ell'}} \int dE' C_{E',\ell m_{\ell}}^{\ell' m_{\ell'}}(E) \psi_{E',\ell m_{\ell}}(r)$ we get
\bea
C_{E',\ell m_{\ell}}^{\ell' m_{\ell'}}(E) \psi_{E',\ell m_{\ell}}(r) &=& \delta(E - E') \delta_{\ell,\ell'} \delta_{m_{\ell},m_{\ell'}}\psi_{E',\ell' m_{\ell'}}^{\rm{reg}}(r)\  \nonumber \\ &-& \sum_{M} A_{JM}^{\ell' m_{\ell'}}(E) \int dr' {\cal K}_{E',\ell m_{\ell}}(r,r')  V_{\ell m_{\ell}; J M}^{\rm{PA}}(r') \phi_{vJ}(r') \eea
Equation (\ref{dss}) is an explicit analytical expression for anisotropic dressed scattering state in partial-wave basis in an ideal loss-less situation. The effect of spontaneous emission may be incorporated phenomenologically by adding spontaneous line width $\gamma_{sp}$ to the stimulated line width  making the total line width $\Gamma_{JM}^{\rm{total}} = \Gamma_{JM} + \gamma_{sp}$. However, spontaneous emission or any other loss process in continuum-bound system can be introduced from first principle through  master equation approach \cite{prl1982gsa,pra1984gsa}.

The analytical expressions for scattering $T$- and other matrices can be analytical derived from asymptotic scattering boundary conditions of (\ref{dss}) as described in Ref \cite{prl2009jisha,jpb2010bdeb}.
The first term ($\psi_{\ell m_{\ell}}^{\rm{reg}}$) of partial-wave ($\ell, m_{\ell}$) scattering state of Eq. (\ref{dss}) is  ground state scattering wave function which is unperturbed by PA coupling while the effect of PA coupling is embodied in the second term through the amplitude $A_{JM}^{\ell' m_{\ell'}}$ where $\ell'$ and $m_{\ell'}$ correspond to the incident or initial  partial-wave of unperturbed state. It is possible that one incident partial-wave would be scattered into another one due to anisotropic nature of laser coupling. The second term describes the effect of two-photon process - one photon exciting the ground continuum into the bound state and another photon de-exciting it back to the same continuum. Since it is a two-photon process, such
effects are possible only in the strong PA coupling regime. As mentioned earlier, at ultracold temperatures only a few low lying partial waves such as $s$- and $p$- waves are predominantly prevalent in the unperturbed continuum. Accordingly, the selection rules of electric dipole transitions allow PA laser
to populate or excite only lowest or low lying rotational level or levels in the excited bound state. For instance, $J=1$ of the bound state can be excited by PA from $s$-wave scattering state. On being de-excited back into the scattering continuum by stimulated transition by the same PA laser, there is a finite
probability of $p$-wave ($\ell=1$) being generated in the continuum \cite{prl2009jisha}.  Here we have restricted our discussion of formalism to one continuum interacting
with only one bound state. However, this formalism can be generalised for other physical situations where two bound states interact with a continuum as shown in \cite{jpb2009agarwal,jpb2009bdeb,jpb2010bdeb}. In such situations where multiple transition pathways interfere quantum mechanically, a host of interesting quantum
effects are expected to arise. For instance,  in V-type bound-continuum-bound atom-molecule interfacial system, vacuum-induced coherence \cite{book:gsa:springer} is shown to arise naturally \cite{daspra}. Furthermore, it is possible to manipulate efficiently  $d$-wave ($\ell=2$) interaction by two lasers driving two PA transitions to two ro-vibrational states \cite{debarxiv}.

\section{4. Quantum interference effects at atom-molecule interface}

Quantum interferences are ubiquitous in physics especially in quantum optics. Quantum interferences between
different atomic transitions has been extensively studied over the years. A host of quantum optical phenomena such as
electromagnetically induced transparency (EIT) \cite{PRL:1989:Harris},  slow light \cite{HauHarris,recentpaperBoyd}, vacuum induced coherence (VIC) in spontaneous emission \cite{book:gsa:springer} etc.  arise due to quantum interference between  transition pathways.
So far spectroscopic quantum interference phenomena are studied mostly with atoms at room temperature or at a temperature higher than that of  ultracold regime. In recent times, quantum interference effect has been  demonstrated in  molecule formation   \cite{atom-molecule} from cold atoms and Autler-Townes splitting \cite{two-photon,moal} in two-photon
PA. Another prominent example of quantum interference is that of Fano \cite{physrev1961fano}.
The hallmark  of  Fano interference is the asymmetric line shape originally observed
in atomic auto-ionisation spectrum.  Eberly \cite{prl1981eberly} described Fano effect as a ``confluence" of coherences. Over the last 50 years, Fano effect has been observed in various physical situations \cite{fanoeffect} although  it has been mostly studied in the context of  atomic auto- and photo-ionisation spectroscopy where the continuum belongs to an excited electronic configuration. In Fano's theory, no spontaneous decay was considered. Agarwal \cite{prl1982gsa,pra1984gsa} introduced spontaneous decay in Fano model  through master equation approach and solved the extended Fano model exactly. With the recent advancement in PA, Fano interference with diatomic ground state continuum has now become important \cite{physreports:2006:shapiro}. The processes that underlie  Fano interference in  PA with MFR and that in an auto-ionosation system  are schematically depicted in Fig.2. 
The asymmetric spectral line shape observed by  Junker {\it et al.} \cite{prl2008hulet} may be an indication of the occurrence of Fano-type quantum interference in PA due to a tunable  MFR \cite{jpb2009agarwal}.

\begin{figure}
\includegraphics[width=4.5in]{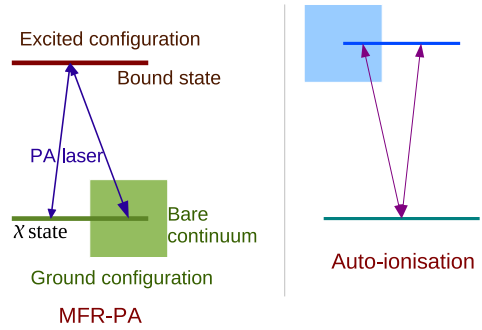}
\caption{A schematic diagram comparing the mechanism of  Fano interference in MFR-modified PA  (MFR-PA) system with
that in auto-ionisation system (see the text). Note that a single laser can drive the continuum-bound and bound-bound transitions meaning that a single photon can be absorbed or emitted through two competing transition pathways.}
 \label{fig1}
  \end{figure}

In the previous section we have presented a method of deriving real-space anisotropic dressed continuum of a system of one excited bound state interacting via electric dipole interaction with the ground continuum. The same method can be applied to obtain dressed continuum of  magnetically induced Feshbach resonance. Let us consider  the simplest two-channel model of MFR. The asymptotic energy (threshold) of the upper (closed) channel is greater than the collision energy $E'$ of the two atoms while the threshold of the lower (open) channel is below the collision energy. The thresholds  of the two channels are tunable by an external magnetic field due to Zeeman effect. The closed channel may support a bound state which may lie very close to the open channel threshold. By varying the magnetic field strength, this bound state can be made to move upward or downward just above or near the open channel threshold. The coupling between this bound state and the open channel scattering state occurs due to hyperfine interaction at an intermediate separation (typically between 20 to 30 Bohr radius).

MFR occurs when the collision energy coincides or nearly coincides  with the binding energy the closed channel quasi-bound state. It follows from the theory of two-body quantum scattering (two-body quantum scattering  is essentially the potential scattering of matter-waves) that when a quasi-bound state is
formed by scattering, scattering phase shift as a function of energy varies rapidly through $\pi/2$ meaning the occurrence of a resonance in scattering amplitude or cross section. Exactly at resonance, the phase shift is $\pi/2$. Since in the Wigner threshold law regime of ultra low energies,  atom-atom scattering is dominated
by $s$-wave only, the total scattering cross section in that regime can be approximated to be the $s$-wave cross section.
Wigner threshold laws state, if the two-body interaction potential is  centrally symmetric and short-ranged meaning the potential has a finite range  or  falls off exponentially at large separation, then in the limit of  energy or  wave number ($k$) going to zero, the partial-wave scattering phase shift $\delta_{\ell} (k)$  behaves as
$\delta_{\ell} (k) \sim k^{2 \ell +1}$. These laws further state that, in case of centrally symmetric
long-ranged potentials with inverse power-law ($\sim 1/r^n$)  where $n>2$) behaviour at asymptotically large separation,
$\delta_{\ell}(k) \sim k^{2 \ell +1}$ for $\ell \le \frac{n-3}{2}$ and
$ \delta_{\ell} (k) \sim k^{n-2}$ for all $\ell >  \frac{n-3}{2}$. For two spherically symmetric ground state $S$ atoms,
the ground state potential $V_g(r)$ goes as $1/r^6$ (van-der Waals type), we have $n=6$. Let us now see how $s$-wave scattering  $a_s$ behaves in the Wigner threshold law regime.  $a_s$ may be  defined from Bethe's expansion formula
\bea
\lim_{k\rightarrow 0} k \cot \delta (k) = - \frac{1}{a_s} + \frac{1}{2} r_e k^2 + \cdots
\eea
where $r_e$ is an effective range of the two-body potential.  The scattering amplitude $f(k)$  is related to $\delta(k)$ by
\bea
f(k) = - \frac{1}{k} \left [ \frac{1}{- \cot \delta(k) + i} \right ] \eea
which, in the limit $k \rightarrow 0$ reduces to \bea f(k) \simeq - \frac{a_s}{1 + i k a_s}. \eea
Thus for two atoms interacting via the potential $V_{g}(r)$, in the limit $k$ going to zero and $|k a_s| <\!< 1$,  $f(k) \simeq - a_s$ becomes a constant. Since two-body effective interaction can be described in terms of scattering amplitude, at ultracold temperatures the effective two-body interaction of cold atoms can be parametric with $a_s$ provided $a_s$ does not diverge. But this is not the case when resonance occurs. At resonance, $a_s$ diverges
and for  $|k a_s| >\!> 1$ we have $f(k) \simeq i/k$. Near resonance, scattering cross section has a Lorenzian shape with the inverse of the Lorenzian width being related to the life time of the two-body resonance complex which we call AMI in case of PA. Thus both PA and MFR have some common features. The differences between the two  are: (i) PA involves electronically excited state, but MFR solely occurs in electronic ground configuration, (ii) PA occurs due to laser interaction and so nature of the continuum-bound coupling is that of electric dipole transition, MFR involves magnetic transitions only, (iii) while MFR can basically tune $s$-wave interaction, OFR is in principle  capable of manipulating not only $s$-wave but also higher partial-wave interactions.  Based on PA, optical Feshbach resonance (OFR) has been developed for tuning $s$-wave scattering length. Unlike MFR,  OFR is found to be an inefficient method as far as tunability of $s$-wave scattering length is concerned.  The inefficiency of OFR mainly stems from the inelastic
loss due to the spontaneous emission from excited bound states. Considering the fact that OFR is applicable for all kinds of atoms - whether they are magnetic or non-magnetic while MFR is applicable only for those atoms which have magnetic moment, it is of interest to devise new methods of suppression of inelastic loss in PA in order to make PA-based OFR
efficient enough for all practical purpose. In attaining that goal, quantum interference at AMI can play an important role as we discuss in what follows next.

For simplicity, let us consider that only $s$-wave part of the ground continuum is coupled the excited ro-vibrational state with $J=1$
by PA in the presence of MFR at ultracold temperature. The dressed state of this system can be written as $\mid E \rangle_{\rm{dr}} = \sum_{M} A_{JM}^{00} \mid v JM \ran + B_E \mid \chi \ran + \int d E' C_{E' \ell=0 \, m_{\ell}=0}^{\ell'=0 \, m_{\ell'}=0}(E) \mid E' \ell=0 \, m_{\ell}=0\ran $ where
 $\mid \chi \ran$ represents closed channel quasi-bound state with $B_E$ being its amplitude coefficient.
 The coefficient $A_{JM}^{00}(E)$  is given by
\bea  A_{JM}^{0 0} = \frac{ (q_{f} + \epsilon)/(\epsilon_B + i)} { \bar{\delta}_L + \epsilon_B \bar{\Gamma}_f/2   - (q_{f} - i)^2/( \epsilon_B + i )
+ i(1 + \bar{\gamma})/2 }  \Lambda^{0  0}_{J M}, \label{a1} \eea
 where $\bar{\delta}_L = (\delta_L - E_{vJ}/\hbar)/\Gamma_{J}^{00}$, $\bar{\Gamma}_f = \Gamma_f/\Gamma_{JM}^{00}$ and $\bar{\gamma} = \gamma_{sp}/\Gamma_{JM}^{00}$   with $\Gamma_{JMJ}^{00} = 2 \pi \mid \Lambda^{0  0}_{J M}\mid^2$ and $\gamma_{sp}$ being the stimulated and spontaneous line width, respectively, of the ro-vibrational state.
 Here $\Gamma_{ f}(E) = 2 \pi \mid \int d r
\psi_{E,00}^{\rm{reg}} (r) V_{\chi}(r)\phi_{\chi}(r)\mid^2 = 2 \pi | V_E |^2 $  is
MFR  line width where $V_{\chi}(r)$ is the interaction potential between the closed  and  open channels
and $ \phi_{\chi}(r) $ is the wave function of the closed channel bound state ($\chi$ state),
\bea
\epsilon_B = \frac{E -
E_{\chi} - E_{\chi}^{\rm{shift}}}{\Gamma_{f}/2} \simeq \bar{E} - \frac{B - B_0}{\Delta k a_{bg} } \eea where $E_{\chi}$ is the binding energy of the $\chi$ state with
\bea
  E_{\chi}^{\rm{shift}} =
 \int d r V_{\chi}(r) \phi_{\chi} (r') \int d r' \rm{Re} {\cal K}_{E, 00}(r,r')
V_{\chi}^*(r')\phi_{\chi} (r') \eea  being  its shift  and  $\bar{E} = E/(\hbar \Gamma_f/2)$ is scaled collision energy, $B$  is the applied magnetic field, $B_0$ is the magnetic field at which Feshbach resonance occurs at zero energy in the absence of PA laser, $\Delta$ is the width (in unit of magnetic field) of
zero crossing in MFR and  $a_{bg}$ is the background scattering length. $\Delta$ is related to Feshbach resonance line width by $\Gamma_f/2 \simeq k a_{bg} \delta \mu \Delta$ where $\delta \mu$ is the difference between the magnetic moment
of the $\chi$ state and the two free atoms in the open channel.
$q_f$ is the well known Fano $q$ parameter which in this context is known as `Feshbach asymmetry' parameter. Explicitly,
  \bea q_f = \frac{V_{\rm{eff}} + \Omega}{\pi
\Lambda_{JM}^{00}(E) V_{E} } \eea
where $\Omega$ is the bound-bound Rabi coupling and
 \bea V_{\rm{eff}} = \rm{Re}  \int d r \phi_{vJ} (r)
V_{JM}^{0 0}(r) \int d r' {\cal K}_{E \ell=0}(r,r')
V_{\chi}(r')\phi_{\chi} (r') \nonumber \eea  is an effective interaction between the two bound states as a result of their
interactions with the  $s$-wave part of the continuum. $V_{JM}^{\ell  m_{\ell}}(r)$,  $\Lambda_{JM}^{\ell m_{\ell}}(E)$ and the Green function
$ {\cal K}_{E,\ell m_{\ell}}(r,r')$
 are defined in section 3.

Let us now analyse how the Eq. (\ref{a1}) can account for Fano profile in MFR-modified PA spectrum. Usually, PA spectrum is obtained by measuring the rate of loss of atoms from  trap due to spontaneous emission from ro-vbrational states. The rate of trap loss at a collision energy $E$ is given by $K_{\rm{loss}}(E) = (\gamma_{sp}/2) \pi  \sum_{M} \sum_{\ell m_{\ell}} \mid A_{JM}^{\ell m_{\ell}}(E) \mid^2 $ which at ultracold energy can be approximated by neglecting the contributions from all $\ell > 0$. PA spectrum is then given by $K_{PA} =  (\gamma_{sp}/2) \lan \pi \mid A_{JM}^{00} \mid^2 \ran$ where $\lan \cdots \ran$ implies thermal averaging over  distribution of relative velocity $v_{rel} = \hbar k/\mu$ which is related to the collision energy by $E = \hbar^2 k^2/2\mu$. Fano minimum in the spectrum is given by the condition $\epsilon_B + q_f = 0$ since $A_{JM}^{00}=0$ for this condition. Assuming $\bar{E} <\!< 1$ (broad Feshbach resonance) we then have $B_{\rm{min}} \simeq B_0 - q_f \Delta k a_{bg}$ where $B_{\rm{min}}$ is the value of $B$ at which Fano minimum occurs. If the ro-vibrational state is long-ranged with its outer turning point lying away
from that of $\chi$ state, then from Franck-Condon principle it follows that we
can neglect the bound-bound Rabi coupling $\Omega$. Note that $q_f$ is independent of laser intensity. However, $q_f$ strongly depends on collision energy. Assuming $V_{JM}^{00}(r)$ to be $r$-independent, $\Lambda_{JM}^{00}(E)$ becomes proportional to the FC factor  $\eta_{vJ} = \int dr \phi_{vJ}(r) \psi_{E,00}^{\rm{reg}} (r)$. At large separation, we have \bea \psi_{E,00}^{\rm{reg}} (r) \simeq \frac{1}{\hbar} \sqrt{\frac{2 \mu }{\pi k}}  \sin [k r + \delta_0^{bg}(k)] \eea
\bea \psi_{E,00}^{\rm{ireg}} (r) \simeq - \frac{1}{\hbar} \sqrt{\frac{2 \mu }{\pi k}}  \cos [k r + \delta_0^{bg}(k) ] \eea
 In the limit $k \rightarrow 0$,  $\delta_0^{bg}$ may behave as  $\delta_0^{bg} \simeq - k a_{bg}$. In case of a broad Feshbach resonance, the background phase-shift $\delta_0^{bg}$ may be finite even  in the $\mu$K energy regime.
This means that for a broad Feshbach resonance, $s$-wave regular scattering wave function in the $\mu$K energy regime can not have a node at long range separation.  In the limit $k \rightarrow 0$, $V_{\rm{eff}}$ becomes energy-independent while
$\eta_{vJ} \propto \sqrt{k} $ and  $V_E  \propto \sqrt{k} $. Therefore, in the limit $k \rightarrow \infty$,  Fano minimum does not depend on $k$. The Fano minimum $B_{\rm{min}}$ crucially depends on $V_{\rm{eff}}$ when $\Omega \simeq 0$. The maximum in Fano profile of PA spectrum as a function of magnetic field depends on a pole of  $A_{JM}^{00}$. There are two poles of  $A_{JM}^{00}$, one of which may lie close to the Fano minimum and the other one away from the minimum. The second pole is responsible for the maximum.

\section{5. Conclusions and outlook}
In conclusion we have combined Fano's method with anisotropic scattering theory to develop an `` anisotropic dressed continuum'' approach for describing the effects of quantum interference between atomic and molecular transitions at low energy. Intensive research efforts are presently going on into forming ultracold ground state molecules in quantum degeneracy limit, in particular polar molecules for their long-range dipolar interactions. Coexistence of atomic and molecular quantum gases will provide an exotic new system for studying many-body quantum physics. Quantum effects at the interface of atomic and molecular states are fundamentally important for quantum control of chemical reactions at low temperatures, and  therefore in the long run, ultracold atom-molecule interface in the quantum regime will relate energy research at a basic level.

\end{document}